# A Turbo Coding System for High Speed Communications[*]

Yan-Xiu Zheng and Yu T. Su[†]

Department of Communication Engineering
National Chiao Tung University
1001 Dar-Shei Road., Hsinchu, 30056
Taiwan

Tel.: +886-3-571-2121   Fax: +886-3-571-0116
non2000.cm88g@nctu.edu.tw   ytsu@mail.nctu.edu.tw

**Abstract**

Conventional turbo codes (CTCs) usually employ a block-oriented interleaving so that each block is separately encoded and decoded. As interleaving and de-interleaving are performed within a block, the message-passing process associated with an iterative decoder is limited to proceed within the corresponding range. This paper presents a new turbo coding scheme that uses a special interleaver structure and a multiple-round early termination test involving both sign check and a CRC code. The new interleaver structure is naturally suited for high speed parallel processing and the resulting coding system offers new design options and tradeoffs that are not available to CTCs. In particular, it becomes possible for the decoder to employ an efficient inter-block collaborative decoding algorithm, passing the information obtained from termination test proved blocks to other unproved blocks. It also becomes important to have a proper decoding schedule. The combined effect is improved performance and reduction in the average decoding delay (whence the required computing power). A memory (storage) management mechanism is included as a critical part of the decoder so as to provide additional design tradeoff between performance and memory size. It is shown that the latter has a modular-like effect in that additional memory units render enhanced performance due not only to less forced early terminations but to possible increases of the interleaving depth. Depending on the decoding schedule, the degree of parallelism and other decoding resources available, the proposed scheme admits a variety of decoder architectures that meet a large range of throughput and performance demands.

---

[*]This work is supported by the National Science Council of Taiwan under Contract 92-2213-E-009-050. Part of this paper was presented at the 15th IEEE PIMRC, Barcelona, Spain, Sep. 2004.

[†]Correspondence addressee.

# 1 Introduction

The performance of iterative decoding of parallel or serial concatenated codes improves as the number of iterations or the interleaving size increases. This is in part due to the fact that the range of the extrinsic information collected for decoding increases accordingly. But the number of iterations and the interleaving size are dominant factors that determine the decoding latency and complexity which, in turn, are often the main concerns in considering the adoptability of such codes in a communication or storage system.

A technique to overcome the dilemma of increasing the range of message exchange and extrinsic information collection or limiting the interleaving size is the recently proposed inter-block permutation interleaver (IBPI) [2] [3]. For a turbo code (TC) using an IBPI, the encoder partitions the incoming data sequence into $L$-bit blocks upon which the IBPI performs intra-block and then inter-block permutations. For example, the IBPI may move contents of a block either to coordinates within the same block or to its $2S$ immediate neighboring blocks so that the IBP-interleaved contents of a block are spread over a range of $2S + 1$ blocks centered at the original block. Such an IBPI is said to have an (left or right) interleaving span $S$.

An in-depth study on the properties and design of IBPI and IBP-interleaved turbo code (IBPTC) is presented in [10]. An example is given in subsection 2.2 to demonstrate the fact that, unlike a conventional turbo code (CTC) that has a fixed range (i.e., the interleaving size) of message (extrinsic information) passing, the range of message passing for an IBPTC increases as the turbo decoder proceeds with further decoding iterations. Moreover, as explained in subsection 2.3, the corresponding decoding latency can be kept at least the same. It suffices to say that, using a fixed block size $L$, a well-designed IBPI with a proper decoding schedule can not only increase the minimum distance of the corresponding IBPTC but also enables an iterative decoder to collect extrinsic information from a range much wider than $L$ and improves the code's performance while maintaining a fixed "local" interleaving size. Note that an IBPI can be build on any existing block-wise interleavers. Using one of them for intra-block permutation, an IBPI has only to add an extra IBP step.

Although IBPTC can provide good bit error rate (BER) performance with relatively small interleaving size and low iteration numbers [10], it is desired that the BER and latency perfor-mance be further improved. It is known that a reliable early termination scheme can accelerate an iterative decoder's decoding speed and reduce the computing power needed for achieving a given performance. The issue of (decoding) termination criterion has been widely discussed [4], [5], [6]. These criteria can be classified into four categories: (i) cross entropy (CE) termination criteria, (ii) sign check (SC) termination criteria, (iii) soft value (SV) termination criteria and (iv) cyclic redundancy check (CRC) termination criteria. The last one guarantees the correctness of decoded bits with a high probability while the others only promise the convergence of the decoded bit sequence. The SC and the CRC termination criteria use the bit operations only while the remaining two categories operate over the floating-point domain. Moreover, CE and SV termination criteria have to optimize threshold for different channel conditions whence is less robust. On the other hand, CRC codes have been widely used in the data link or higher layer as part of the error-control mechanism and is an indispensable component of a packet-oriented data communication system. Using CRC codes as a part of the termination criterion thus causes little or no extra complexity.

Since an IBPI permutes bits in a block to neighboring blocks within its span and CRC code detection and other early termination schemes are performed in a block-by-block fashion, termination time variation over different blocks is inevitable. On the other hand, the special structure of IBPIs implies that bits in neighboring blocks are algebraically related, hence the information about bits in terminated blocks can be used to help decoding bits in unterminated blocks, i.e., one can actually take advantage of the termination time variation if proper statistical information can be extracted from terminated blocks.

This paper presents a new novel codec structure that allows collaborate decoding among different blocks in the above-mentioned sense. When used in conjunction with a highly reliable multiple round termination test using both SC and a CRC code, the proposed coding scheme yields low latency and requires small average decoding iterations while achieving very impressive performance.

We begin our presentation in the next section with a description of the proposed codec system model, iterative decoding procedure and discussion on the latency issue. In Section 3, we present the structure of a variable termination time IBPTC decoder. Section 4 describes multiple-round termination criteria and tests and the following section addresses the issue of storage requirement and suggests a general dynamic memory management algorithm. Section 6 provides some numerical examples that validate the superiority of the proposed coding scheme and finally, in Section 7 we summarize our main results.

## 2 System Model and IBP Behavior

### 2.1 System model

Shown in Fig. 1 (a) is a generic block diagram for a communication system using an IBPTC. The input data sequence $\mathbf{X}$ is partitioned into blocks of the same length, $\{\mathbf{X}_1, \mathbf{X}_2, \cdots\}$, where $\mathbf{X}_i$ is a row vector of length $L - K_{CRC}$ representing the $i$th block. They are CRC-encoded into $\mathbf{W} = \{\mathbf{W}_1, \mathbf{W}_2, \cdots\}$, where $\mathbf{W}_i = \{w_{i0}, w_{i1}, \cdots, w_{i(L-1)}\}$ is a row vector with length $L$. $\mathbf{W}$ is formed by padding at the end of each data block parity bits that are the coefficients of the remainder polynomial $r(x)$ obtained by dividing a data polynomial associated with a data block by a binary generator polynomial $g(x)$ of order $K_{CRC}$. Of course, the degree of $r(x)$ is less than $K_{CRC}$. The corresponding probability of undetectable error is roughly equal to $2^{-K_{CRC}}$. In other words, longer CRC codes possess better error detection capability.

The CRC encoder output $\mathbf{W}$ and its IBPI-permuted version $\mathbf{W}' = \{\mathbf{W}'_1, \mathbf{W}'_2, \cdots, \}$ are then encoded to form the coded sequence $\mathbf{Z} = \{\mathbf{Z}^1, \mathbf{Z}^2, \mathbf{Z}^3\}$, where $\mathbf{Z}^i = \{\mathbf{Z}^i_1, \mathbf{Z}^i_2, \cdots\}$ and the superscript $i$ is used to denote the systematic part ($i = 1$), the first encoder's output (parity) sequence ($i = 2$) and the second encoder's output sequence ($i = 3$). Fig. 1 (b) shows the structure of a TC encoder where the only difference between a CTC and an IBPTC encoders is the interleaver used.

The receiver uses one or multiple APP decoding unit (ADU) like that shown in Fig. 1 (c) to decode the corresponding received baseband sequence $\mathbf{Y} = \{\mathbf{Y}^1_1, \mathbf{Y}^2_1, \mathbf{Y}^3_1, \mathbf{Y}^1_2, \mathbf{Y}^2_2, \mathbf{Y}^3_2, \cdots\}$, where $\mathbf{Y}^i_j$ is the subsequence corresponding to $\mathbf{Z}^i_j$; other notations are defined in the next section. An ADU consists of an APP decoder and a termination condition checker (TCC). It also performs the corresponding interleaving or de-interleaving and other related operations but for simplicity we do not show these operations in the figure.

The TCC applies CRC check and/or other forms of termination tests (TTs) to verify if the APP decoder output satisfies the termination criterion. An affirmative answer leads to the decision to stop decoding (terminate) the block in question and this is the only possible early termination (ET) opportunity for CTCs. Besides such regular ETs, however, there are two other ET opportunities for IBPTCs since no matter whether the decoder output passes the TT, the corresponding soft output is interleaved or de-interleaved to the neighboring blocks within the IBP span. The TCC will then examine each block within the span and check if the content has been filled with terminated decisions. If such a block is found the TCC will issue a termination decision accordingly. The TCC can also run TTs on these neighboring blocks and make a termination decision. We refer to the latter two ET possibilities as extended ETs. Note that a decoding iteration consists of two DRs that are respectively responsible for decoding the pre-permuted (non-interleaved) $\mathbf{Z}_j^2$ and post-permuted (interleaved) blocks $\mathbf{Z}_j^3$ and CRC check is feasible for pre-permuted blocks only. Hence in the first DR one can perform both regular and extended ETs, but in the second DR, only extended ET is viable unless the TT does not involve a CRC check. Examples are given in Section 5 to further elaborate this property of IBPTCs.

## 2.2 Graphical representation of an IBPTC

Fig. 2 is a graphical representation for the system of Fig. 1 with a symmetric IBPI of interleaving span $S = 1$ and an input data sequence $\mathbf{X}$ of five-block duration. The black-, red-, green- and blue-colored squares represent respectively the functions of convolution codec, CRC codec, IBPI and the channel effect. This extended graph is used to describe the iterated decoding behavior. As one can see, the content of a given block, say $\mathbf{W}_2$, is interleaved to parts of itself $\mathbf{W}_2'$ and the two neighboring blocks $\mathbf{W}_1'$ and $\mathbf{W}_3'$ to its immediate left and right.

At the first DR, the decoder uses $\mathbf{Y}_i^1$, $\mathbf{Y}_i^2$ to decode block $\mathbf{W}_i$. The extrinsic information of, say $\mathbf{W}_1$ and $\mathbf{W}_5$, is interleaved for use in decoding $\mathbf{W}_1'$, $\mathbf{W}_2'$ and $\mathbf{W}_4'$, $\mathbf{W}_5'$ in the second DR. It is easy to see that, in decoding the third block $\mathbf{W}_3$ at the beginning of the second iteration, the decoder can use as a priori information some message passed from all five neighboring blocks. In general, an IBPTC decoder can exploit information collected from $4SI + 1$ adjacent blocks in $I$ iterations while, as will be shown in the ensuing subsection, the average decoding delay between two output blocks is kept fixed. This massage passing range expansion capability implies that an IBP interleaver can have an unbounded equivalent interleaving depth (size) that is constrained only by the numbers of turbo decoding iterations and the data blocks involved in decoding while keeping the interleaving delay per iteration bounded by its local interleaving depth.

## 2.3 The latency issue

Although the interleaving process of an IBPI is defined by the composition of the intra- and inter-block permutations, it can be implemented by a single step. The encoder knows to which position each bit (or sample) in a given block should be moved and can do so immediately after it receives each incoming bit. But to encode a given, say the $i$th, interleaved block $\mathbf{U}'_i$ into $\mathbf{X}^2_i$, it has to wait until the complete $(i+S)$th block is received. The time elapsed between the instant the encoder receives the first bit of the $i$th block and the moment when it receives the last bit of the $(i+S)$th block and outputs its first encoded bit of $\mathbf{X}^2_i$ is simply $(1+S)L$-bit durations. By contrast, a CTC with a block size of $L$ bits has an encoding delay of approximately $L$ bits.

The single-round interleaving (or de-interleaving) delay (SRID) is proportional to the encoding delay. But the decoding delay is a much more complicated issue. For the first decoding of each incoming block, there can be zero waiting time, but for later DRs the corresponding delays depend on, among other things, the decoding schedule used. With the same block size, the decoding delay of the first received block for the CTC is definitely shorter than that for the IBPTC. But if one considers a period that consists of multiple blocks (otherwise one will not have enough blocks to perform inter-block permutation) and takes the decoding schedule into account, then the average decoding latency difference can be completely eliminated. This is because the APP decoder (including the interleaver and deinterleaver) will not stay idle until all blocks within the span of a given block are received. Instead, the APP decoder will perform decoding-interleaving or deinterleaving operations for other blocks according to a predetermined decoding schedule before it can do so for the given block (and the given DR).

If we define the total decoding delay (TDD) as the time span between the instant a decoder receives the first input sample (from the input buffer) and the moment when it outputs its last decision then both the IBP and classic approaches yield the same TDD even if only one APP decoder is used. We use the following example and Fig. 3 to support our claim; its generalization is straightforward.

Suppose we receive a total of 7 blocks of samples (in a packet, say) and want to finish decoding in 2 iterations (4 DRs). One can easily see from Fig. 3 that a classic TC decoder would output the first decoded block in 4 $DT$ cycles, where $DT$ is the number of cycles needed to perform a single-block APP decoding plus SRID. The IBPTC decoder, on the other hand, needs 10 DT cycles to output its first decoded block. However, if one further examines the decoding delays associated with the remaining blocks, then one finds they are 8, 12, 16, 20, 24, and 28 DT cycles for the CTC decoder while those for the IBPTC decoder are 14, 18, 22, 25, 27 and 28 DT cycles, respectively. So in the end, both approaches reach the final decision at the same time.

It can be shown that, for a decoder with $2N$ DRs and $S = 1$, both decoders result in a constant delay of $2N$ DT cycles between two adjacent output blocks, except for the first block and the last $2N - 1$ blocks. For an $S = 1$ IBPTC, the decoder requires a first-block decoding delay (FBDD) of $N(1 + 2N)$ DT cycles while the FBDD for the classic TC is only $2N$ DT cycles. The inter-block decoding delays (IBDD, i.e., decoding latency between two consecutive output blocks) for the last $2N - 1$ output blocks of an IBPTC decoder using a decoding schedule similar to that shown in Fig. 3 (e.g., the one shown in Fig. 5 (a)) form a monotonic decreasing arithmetic sequence $\{2N - 1, 2N - 2, \cdots, 1\}$ (in DT cycles). The IBDD of a CTC decoder remains a constant $2N$ DT cycles. On the average, both codes give the same IBDD.

The above assessment on the encoding/decoding delay is made under the assumptions that both codes use the same block size $L$, no early termination mechanism is applied, and a single APP decoder is used. As was mentioned in the previous subsection, an IBPTC has an equivalent interleaving depth that grows as the number of iterations increases, hence [10] with an identical block size an IBPTC always outperforms its classic counterpart. In other words, an IBPTC requires a smaller block size and thus less decoding delay to achieve the same performance. The IBPTC is also naturally suited for parallel decoding. An example of decoding an IBPTC with multiple APP decoders is given in Section 5; see Fig. 5 (a). Although both IBPTCs and CTCs can use multiple decoders for parallel decoding and apply an early termination method to shorten the decoding latency, we will explain in the next section and prove numerically in Section 6 that the former class does derive much more benefits.

Before proceeding to the main discourse, it is worthwhile to recapture an alternative viewpoint on the concepts of IBP based on the the above example. For a CTC with block size of $7L$ bits, the FBDD is 28 DT cycles. But if one divides a $7L$-block into 7 subblocks and a special interleaver which performs successive intra-subblock and inter-block permutations on these subblocks, the corresponding decoding delays in DT cycles for these subblocks are 14, 18, 22, 25, 27, and 28, respectively. Therefore, although both code structures result in identical TDD the IBPTC structure is able to supply partial decoded outputs much earlier. This feature, when combined with suitable decoding schedule and implementation resources, become very beneficial for high throughput applications.

## 3 Iterative Decoder with Variable Termination Time

A conventional iterative decoder is composed of two or more APP decoders that will not stop decoding until a fixed number of decoding iterations have been performed. With an early-termination mechanism in place, as shown in Fig. 1 (a), the decoding procedure can stop at the end of an iteration (two DRs) or at the end of a DR. We will refer to such a decoder as variable termination time APP (VTT-APP) decoder. All termination tests (TTs), whether they are used in CTCs or IBPTCs, incur additional computational complexity, which is usually more than compensated for by the reduced average DRs brought about by the use of a TT.

### 3.1 Generation of extrinsic output with a termination test

Let $\Lambda(w) = \log \frac{p(w=+1)}{p(w=-1)}$ be the log-likelihood ratio of the random variable $w$ where $p(\cdot)$ denotes the probability density function of $w$. If $w_{jk}$ represents the $k$th bit of the $j$th block and $\Lambda^{(i)}(w_{jk})$, $\Lambda_e^{(i)}(w_{jk})$ denote the corresponding estimated log-likelihood ratio and the extrinsic information at the $i$th APP DR's output, we have [4]

$$\Lambda_e^{(i)}(w_{jk}) = \Lambda^{(i)}(w_{jk}) - \Lambda_e^{(i-1)}(w_{jk}) - L_c \cdot y_{jk}^1, \qquad (1)$$

We assume that $\Lambda_e^{(-1)}(w_{jk}) = 0$, $\forall j, k$. $L_c = 4aE_s/N_0$ represents the channel reliability, where $a$ is the signal amplitude which is usually normalized to 1 for additive white Gaussian noise (AWGN) channel, $E_s$ being the signal energy per symbol while $N_0$ is the noise power spectral density.

The $i$th VTT-APP decoder has the received baseband vectors $\mathbf{Y}^1, \mathbf{Y}^2, \mathbf{Y}^3$ and the a priori information $\{\Lambda_e^{(i-1)}(w_{jk})\}_{k=0}^{k=L-1}$ as its input and outputs $\{\Lambda_e^{(i)}(w_{jk})\}_{k=0}^{k=L-1}$ by using (1) to the $(i+1)$th decoder as a priori information until $i = D_{max}$, where $D_{max}$ is the maximum allowed APP DRs; see Fig. 1 (c).

A tentative decision $\hat{w}_{jk}^i$ on the $k$th bit of the $j$th block at the end of the $i$th APP DR can be obtained by

$$\hat{w}_{jk}^i = \begin{cases} 0 & , \Lambda^{(i)}(w_{jk}) \geq 0, \\ 1 & , \Lambda^{(i)}(w_{jk}) < 0. \end{cases} \tag{2}$$

Let $Q(\widehat{W})$ be the termination indicator for the tentative decision vector of the $j$th block at the $i$th DR, $\widehat{\mathbf{W}}_j^i = \{\hat{w}_{j0}^i, \hat{w}_{j1}^i, \cdots, \hat{w}_{j(L-1)}^i\}$, $0 < i \leq D_{max}$,

$$Q(\widehat{W}_j^i) = \begin{cases} 1, & \text{if } \widehat{W}_j^i \text{ satisfies the termination condition} \\ 0, & \text{else} \end{cases} \tag{3}$$

Then the conditional soft value $\Lambda_S^{(i)}(w_{jk})$ and the extrinsic information $\Lambda_{e,S}^{(i)}(w_{jk})$ are given by

$$\Lambda_S^{(i)}(w_{jk}) = \begin{cases} \log \frac{P(w_{jk}=+1|Q(\widehat{W}_j^i)=1)}{P(w_{jk}=-1|Q(\widehat{W}_j^i)=1)} & , Q(\widehat{W}_j^i) = 1 \\ \Lambda^{(i)}(w_{jk}) & , Q(\widehat{W}_j^i) = 0 \end{cases} \tag{4}$$

and

$$\Lambda_{e,S}^{(i)}(w_{jk}) = \Lambda_S^{(i)}(w_{jk}) - \Lambda^{(i)}(w_{j,k}). \tag{5}$$

The extrinsic information $\Lambda_{e,V-A}^{(i)}(w_{jk})$ of a VTT-APP decoder then becomes

$$\Lambda_{e,V-A}^{(i)}(w_{jk}) = \Lambda_{e,S}^{(i)}(w_{jk}) + \Lambda_e^{(i)}(w_{jk}) = \Lambda_S^{(i)}(w_{jk}) - \Lambda_e^{(i-1)}(w_{jk}) - L_c \cdot y_{jk}. \tag{6}$$

When $Q(\widehat{W}_j^i) = 0$, we have

$$\Lambda_{e,V-A}^{(i)}(w_{jk}) = \Lambda_e^{(i)}(w_{jk}) = \Lambda^{(i)}(w_{jk}) - \Lambda_e^{(i-1)}(w_{jk}) - L_c y_{jk}. \tag{7}$$

which is consistent with the conventional iterative decoding algorithm. The extrinsic information of the VTT-APP decoder can be further formulated as

$$\Lambda_{e,V-A}^{(i)}(w_{jk}) = \Lambda_S^{(i)}(w_{jk}) - \Lambda_{e,V-A}^{(i-1)}(w_{jk}) - L_c y_{jk}. \tag{8}$$

The resulting VTT-APP decoder is shown in Fig. 1 (c).

To ease the burden of computing the conditional log-likelihood function that appears in (4), we make the idealized assumption that the termination test is perfect, i.e.,

$$P(\hat{w}_{jk}^i \text{ is correct}|Q(\widehat{W}_j^i) = 1) = 1, \ \forall \ k$$

With this perfect termination test assumption, (4) becomes

$$\Lambda_S^{(i)}(w_{jk}) = \begin{cases} \Lambda^{(i)}(w_{jk}) \cdot \infty, & Q(\widehat{W}_j^i) = 1 \\ \Lambda^{(i)}(w_{jk}), & Q(\widehat{W}_j^i) = 0 \end{cases}, \quad (9)$$

and (8) can be rewritten as

$$\Lambda_{e,V-A}^{(i)}(w_{jk}) = \begin{cases} \Lambda^{(i)}(w_{jk}) \cdot \infty, & Q(\widehat{W}_j^i) = 1 \\ \Lambda_e^{(i)}(w_{jk}), & Q(\widehat{W}_j^i) = 0 \end{cases}. \quad (10)$$

The perfect termination assumption actually makes the computation of extrinsic information or soft output easier as when the tentative decision vector $\widehat{W}_j^i$ meets the termination condition, then $\Lambda_{e,V-A}^{(i)}(w_{jk})$ has only two values $\pm\infty$. A practical approximation is to assign a fixed large number to $\Lambda_{e,V-A}^{(i)}(w_{jk})$. However, it should be noted that, after interleaving or de-interleaving, the large metric value will be passed to neighboring blocks and then to the corresponding partial path metric computers, eliminating other branches which are not associated with these bits. Hence the passing of the extrinsic information of these perfect detected bits to neighboring blocks further reduce the complexity of the associated APP decoder. Moreover, as the APP decoder selects survivor branches based on the relative magnitudes of the partial path metrics only, the actual value assigned to $\Lambda_{e,V-A}^{(i)}(w_{jk})$ is immaterial. In fact, it can be as simple as a binary sign telling the APP decoder which branches should be eliminated.

All these nice features depend, besides the IBP design, on the assumption of a good block TT which is the subject of the following section. Note that although there is no perfect TT and the probability that a TT gives a wrong block termination decision is nonzero, an incorrect block termination yields only a few erroneous soft bits and they are spread to only parts of the neighboring blocks after IBP interleaving or deinterleaving. The influence of these wrong indications will be diluted in subsequent DRs and the simplified soft value computing formulae of the last two equations result in no catastrophic failure as our numerical results will demonstrate later.

# 4 Multiple-Round Termination Tests

[6] summarized various TTs for TCs using sign check, soft values and CRC checks. The sign check termination test (SCTT) compares the tentative decoded bits from two successive rounds. A tentative decoded block passes the test if most or all of them are consistent. The soft value termination test (SVTT) compares the soft value(s) with a threshold; the soft values can be the reliability of tentative decoded soft bits, the average soft value of a block, the extrinsic value of the least reliable bit etc. The CRC termination test (CRCTT) uses the CRC result to decide if further decoding of a block is needed. SCTT and CRCTT operate over bit level but SVTT operates over the real domain. The performance of SVTT is subject to the choice of the threshold which, in turn, is a function of the channel condition and code structure. Moreover, the convergence rate of soft bit values also depends on the above two factors [6, 11]. In short, the classes of CRCTT, SCTT or their variations have the complexity and robustness advantages over the class of SVTT.

## 4.1 A general algorithm

It is well known that a statistical decision based on a single observation is inferior to that based on multiple observations which, however, require a longer observation time (or equivalently, larger sample size). Most of the termination criteria are, in a sense, based on a single-round test. They either compare or manipulate some values corresponding to two consecutive DRs, or just check a single DR output to make a stop-or-continue decoding decision. Our proposed termination criteria are based on multiple-round tests (MRT). The MRT has the distinct capability of balancing performance (reliability of the test) and cost (time or sample size needed to make a decision). The rationale of such a test is similar to that of the variable dwell time PN acquisition scheme [7]. A dismissal on a decoder output is issued as soon as it fails a single test but a decision to stop decoding a block has to wait until the same block is verified by several rounds of test. Therefore, incorrect tentative decoder outputs are quickly discarded while any final decision on a block is prudently made. Since most of the DRs do not lead to the final decision of a block, an MRT that consists of a series of simple, short-duration tests spends much less time and overhead on these intermediate DRs than those required by a single long-duration test. Furthermore, the additional verification test rounds greatly reduce the probability of false terminations and give an MRT more robust and reliable decisions, avoiding spreading incorrect information to neighboring blocks. Moreover, since a correct termination on a certain block helps bringing earlier terminations to its adjacent blocks, the average decoding latency is shortened as well.

A flow chart of the general multiple-round termination test (MRTT) is shown in Fig. 4. In this figure, $i$ is used to denote the $i$th DR, $p$ represents the number of times a block has passed the test and can be regarded as a quality indicator, $m$ is the required quality condition and $D_{max}$ is the maximum number of DRs allowed. Either $p = m$ or $i = D_{max}$ will force the decoding process to be terminated. As discussed in subsections 2.1 and 2.2, a TT is performed at the end of an iteration (even DRs) or the beginning of an odd DR. For the latter case, a TT means checking if all pre-permuted blocks within its span have satisfied the termination condition. A special case of MRTT is the multiple-round SCTT of [6]. It was found that the bit error rate (BER) performance improves as the number of test rounds increases.

As mentioned before, we shall not consider the class of SVTTs. The following subsections describe multiple-round CRCTT, SCTT and a hybrid CRC-SC TT.

## 4.2 T1.m: the $m$-round CRCTT

This scheme is based on an $m$-round CRC test. A block is said to pass the $m$-round CRCTT if all $m$ consecutive tentative decision vectors $\widehat{\mathbf{W}}_j^{i-m+1}, \widehat{\mathbf{W}}_j^{i-m+2}, \cdots, \widehat{\mathbf{W}}_j^i$ succeed in passing the same CRC test, i.e., $I_{CRC}(\widehat{\mathbf{W}}_j^l) = 1$, $l = i - m + 1, i - m + 2, \cdots, i$ and $i \leq D_{max}$, where

$$I_{CRC}(\widehat{\mathbf{W}}) = \begin{cases} 1, & \widehat{\mathbf{W}} \text{ passes CRC condition} \\ 0, & \text{otherwise} \end{cases}. \tag{11}$$

As the error detection capability of a CRC code is an increasing function of the code length, one can trade the order $m$ for the code length.

## 4.3 T2.m: the $m$-round SCTT

This TT [6] compares tentative decoded bits in $m$ ($m \geq 2$) consecutive DRs or iterations. The decoder stops when the $n$th tentative decision vector, $i \leq D_{max}$, are the same with the previous $m - 1$ tentative decision vectors, i.e.,

$$\hat{w}_{jk}^{i-m+1} = \hat{w}_{jk}^{i-m+2} = \cdots = \hat{w}_{jk}^i, \forall\ k,\ 0 \leq k < L. \tag{12}$$

Note that MR-SCTT checks the convergence of tentative decisions, it does not guarantee the convergence to the correct decisions.

## 4.4  T3.m: the $m$-round hybrid termination test (MR-HTT)

Unlike CTCs, errors in TTs for IBPTC will propagate to different blocks and might lead to a catastrophic consequence. A highly reliable TT can be obtained by increasing $m$ or it can be obtained by incorporating multiple criteria in a single round. A block that passes both CRC and SC tests is more reliable than one that passes only a single test.

Hence, we suggest the hybrid termination criterion

$$I_{CRC}(\widehat{W}_j^l) = 1, \ \forall \ l, i - m < l \leq i, \ i \leq D_{max}, \tag{13}$$

and

$$\hat{w}_{jk}^{i-m+1} = \hat{w}_{jk}^{i-m+2} = \cdots = \hat{w}_{jk}^i, \forall \ k, 0 \leq k < L, \ i \leq D_{max}. \tag{14}$$

If the CRC-8 is used, the undetect error probability is approximately $2^{-8}$ only. The probability that the sign check does not match the CRC result is of the order $2^{-16}$ or $2 \times 10^{-5}$. Using a longer CRC code increases the reliability of a CRC test but it also implies an increase in the overhead. Additional sign consistency check is the price we paid for using a short CRC code to cut down the CRC overhead.

## 4.5  Genie termination test

Genie TT is a hypothetic ideal test that is capable of verifying the tentative decision vector without error. The performance of this ideal test is used as the ultimate bound for reference purpose.

At the first glance, we might expect the hybrid test or higher-order (larger $m$) tests to take more DRs since a received block is less likely to pass both SC and CRC or a higher-order requirement. But the fact is that a correct block decision, through the IBP interleaving, will help other blocks to meet the termination condition sooner while an incorrect one tends to has an adverse effect. Our numerical experiment indicates that the hybrid test not only gives better performance but also requires less DRs. This is another advantage of IBPTCs that is not shared by CTCs.

# 5 VTT IBPTC Decoder with Finite Memory Space

## 5.1 Decoding schedule and early terminations

In subsection 2.3, we have demonstrated the importance of the decoding schedule in minimizing the decoding delay. Parallel decoding is a popular design option to shorten the latency. Fig. 5 (a) shows a multiple expanding-window zigzag scheduling table for decoding an IBPTC with interleaver span $S = 1$ and four ADUs, denoted respectively by **a**, **b**, **c** and **d**. Data blocks processed in the odd rows are in the original (pre-permutation) order while those processed in the even rows are in the interleaved (post-permutation) order. Each dashed or dotted zigzag curve represents the schedule for an ADU. The symbol $\mathbf{x}_{mn}$ denotes the $n$th DR of the $m$th phase in the ADU **x**'s schedule, where a DR represents the APP decoding of a pre- or post-permuted block and the associated interleaving or de-interleaving and the $m$th phase refers to the $m$th parallel line associated with a schedule. Obviously, the $m$th decoding phase of **x** is followed by the $(m+1)$th decoding phase to its right.

Take ADU **c**'s decoding schedule as an example, its first DR of the first phase $\mathbf{c}_{11}$ corresponds to the first DR of Block 3 while the first phase' second DR $\mathbf{c}_{12}$ corresponds to the second DR of Block 2. $\mathbf{c}_{12}$ can be performed as the scheduling table shows that Blocks 1,2,3 have been decoded once and the corresponding extrinsic information output has been inter-block interleaved so that the post-permuted Block 2 is completely filled and is ready for a new DR. **c** finishes its first phase after $\mathbf{c}_{13}$ is done. It then proceeds with the first DR of the next phase $\mathbf{c}_{21}$, i.e., the first DR of Block 7. An ADU can not start a new DR until the DR on its left is completed, e.g., $\mathbf{a}_{2k}$, $k > 2$ cannot start unless the DR corresponding to $\mathbf{d}_{12}$ is finished.

As mentioned at the end of subsection 2.1, an ADU can make both regular and extended early termination decisions (ETDs) in odd rows' DRs only unless a non-CRC-based TT is used. For DRs in even rows, however, extended ETDs are still feasible. For example, in $\mathbf{a}_{23}$ ( $\mathbf{c}_{25}$) we check if block 3 passes the TT and ET on this block becomes effective if affirmative. ADU **a** (**c**) then go on to examine whether $\mathbf{a}_{24}$ ($\mathbf{c}_{26}$) is necessary by checking whether both $\mathbf{c}_{13}$ and $\mathbf{d}_{13}$ ($\mathbf{a}_{25}$ and $\mathbf{b}_{25}$) pass the TT as well. When this condition is satisfied, decoding of block 2 is terminated. On the other hand, in $\mathbf{b}_{24}$ no TT is performed but after de-interleaving its output we run a TT on the content of $\mathbf{b}_{25}$, which contains de-interleaved outputs from $\mathbf{d}_{14}$ and $\mathbf{a}_{24}$. Block 2 is terminated and $\mathbf{b}_{25}$ is no longer needed (because of our schedule and the IBP structure, $\mathbf{c}_{25}$ and $\mathbf{d}_{25}$ can not yet be verified although extrinsic information from $\mathbf{b}_{24}$ will be passed on to them) if the TT result is positive.

## 5.2   Memory management

From the above discussion, it is clear that decoding more than 10 blocks at the same time requires no small storage area for ASIC or DSP implementation. One should therefore try to make the most of the memory space available. The decoder needs space to store (I) received samples undergoing decoding, (II) extrinsic information, (III) decoded bits to be forwarded to a higher layer for further processing, and (IV) received samples awaiting decoding. The management of the last category, assuming no buffer overflow, requires only an indicator signal to forward a new block of received samples to the part of the storage area designated for category (I) that was just released due to a termination decision.

Category (III) is needed because of the termination variation across blocks. Its management is straightforward and, besides, it requires much less storage space. As mentioned in Section 3.1, assigning the extrinsic values for TT-approved bits a constant large value is equivalent to using a (special) binary-valued bit to indicate which partial paths should survive in the APP decoding process. Hence the decoded bits serve the dual purposes of representing the decoder decisions and bookkeeping the survivor paths. The management of categories (I) and (II), however, needs more efforts and careful considerations.

As long as the probability of termination-defying blocks exists, practical latency consideration will force us to set an upper limit $D_{max}$ on the number of DRs. It can be shown that an unterminated block will prevent the decoder from discarding $\mathbf{Y}^3$ associated with those terminated blocks within a $2S$ neighborhood. When the number of blocks that terminate at or around the $D_{max}$th DR is large so will be the memory required. Hardware constraint thus imposes another threshold $M_{max}$, the maximum affordable (allowable) memory units (MU) where an MU refers to the space for storing categories (I) and (II) associated with a block of data in the decoder. As our sole purpose is to demonstrate the critical role a memory manager plays in the VTT-APP decoder, we assume, for simplicity, that the same number of bits is used to represent the extrinsic information of a bit and the corresponding received sample. An MU is thus assumed to contain $KL$ bits, where a $K$-bit word is used to store either the extrinsic information or received baseband sample associated with a transmitted bit.

Because of the termination time variation nature of our decoder, a memory manager has to take into account both thresholds, $D_{max}$ and $M_{max}$ so as to optimize the performance. When a block has failed to pass the TT for $D_{max}$ times, it will automatically be discarded and the MUs storing the corresponding categories (I) and (II) information are released accordingly. Chances are more than one block that reach the threshold $D_{max}$ simultaneously and it is even more likely that the decoder runs out of MUs before a block reaches the threshold $D_{max}$. For both cases, one should then give up decoding one or some of the unterminated blocks. It is both reasonable and intuitively-appearing to terminate the most ancient block, i.e., the one which has failed the TT most often. To distinguish from the regular and extended ETs described in subsection 2.1, we refer to these memory shortage induced terminations as forced early terminations.

Fig. 5 (b) shows a finite-memory IBPTC decoding procedure for one phase of an ADU. The procedure involves APP decoding, interleaving, deinterleaving, regular and extended ETDs, and memory check and release. The last two operations are collectively called the memory management scheme which is responsible for verifying if there is enough memory during the decoding process and make a proper memory-release decision if there is not enough storage space. As there is no computing involved at all, the complexity is moderate at most.

Denote by $M_F$, $M_d$ and $M_R$, the numbers of free (unused) MUs, ADUs, and the required MUs for storing one received block. It follows that $M_R = x$ for a rate $R = 1/x$ turbo code. The decoder is initialized with $M_F = M_{max}$. An ADU begins a phase by checking if $M_F < M_R$ (Box 2) where the additional MU is for storing extrinsic information. If $M_F$ does not meet the condition, the memory manager determines which block is to be discarded, makes a forced ETD, and releases the related storage space (Box 4). Otherwise, the decoder moves the received samples of the new block from where they were saved (in the buffer area) to the corresponding category (I) MUs (Box 3).

Deciding which block is to be given up is simple and clear since our decoding schedule allows only a single most ancient block in its left-most active column at any time. When a forced ETD is made the ADU makes hard decisions on the block to be discarded and releases the related categories (I) and (II) MUs. As the discarded block is always the most ancient block and our decoding schedule is such that all blocks to its left must have been terminated for one reason or another, we are left with the problem of dealing with the $S$ adjacent blocks to its right if they have not been terminated. At least two alternatives exist for solving this problem. The first solution, which leads to better performance at the cost of higher complexity, is to interleave or de-interleave the extrinsic values for use in decoding the $S$ blocks to its right without further updates. The second one is to make hard-decisions (stop any further decoding) on all $S$ blocks within its (right) span, releasing their category (I) MUs while keeping their category (II) MUs for use in decoding other related blocks.

At beginning of each DR, we ask the decoder whether the scheduled DR is needed (Box 5). Unless the decoder has been notified to by-pass the ensuing DR, we still have to ask if the space for storing the extrinsic information of the coming DR is available. When such a space is not available ($M_F = 0$) the decoder has to find room for the next DR by discarding the most ancient unterminated block and following the memory release procedure described above (Box 7). The operations in Box 9 include those described in the last paragraph of subsection 2.1. When the TCC makes a regular or extended ETD, the memory manager releases the corresponding category (II) and parts of category (I) memory (Box 10), memory release procedure 1) and notifies the decoder that further decodings on these blocks are no longer necessary.

# 6 Numerical Examples

The simulation results reported in this section is based on the following assumptions and parameters. The component code of the rate=1/3 TC, $G(D) = [1, 1 + D^2 + D^3/1 + D + D^3]$, and the CRC-8(= "110011011") code used are the same as those specified in the 3GPP standard [8]. The APP decoder uses the Log-MAP algorithm and the IBP algorithm of [3] while the interleaving length and span are left as variables; $M_R = 3$ MUs and $N = 1000$ per computer run are assumed. Except for the Genie TT, our simulations do not assume perfect block termination.

The effects of various TTs on the IBPTC VTT-APP decoder performance for the system with $S = 1$, $L = 400$, $D_{max} = 30$ and (zero-)tail-padding encoding are shown in Figs. 6 ∼ 7. Multiple-round CRCTT, SCTT and HTT are considered. For comparison, we include performance curves of the decoder using the genie TT, that with fixed 20 and 30 DRs (10 and 15 iterations) and that of the CTC with block length $L = 800$ using the genie TT with $D_{max} = 30$.

BER performance improves as the number of test rounds $m$ increases no matter which TT is used. Fig. 6 shows that T1.3 outperforms T1.2 for $E_b/N_0$ greater than 0.3 dB. Tests using sign-check alone, T2.3 and T2.5, are inferior to other termination criteria since, as mentioned before, the class of sign-check tests check if decoded bits converge but can not guarantee the quality of the tentative decoded vectors. Incorrect termination decisions will spread false information to the neighboring blocks through interleaving and result in degraded performance. T1.3, T3.2, T3.3 and the one with fixed 30 DRs yield the best performance and they are almost as good as the genie TT. Using T3.2 for early termination, the IBPTC has $0.4 \sim 0.6$ dB gain against the CTC for BER=$10^{-5} \sim 10^{-6}$ although the average decoding delay per DR for both codes are about the same.

Fig. 7 shows the average DR performance of various TTs. Except for the two sign-check tests, all TTs require less than 20 or 10 APP DRs (10 or 5 iterations) when $E_b/N_0$ is greater than 0.2 or 0.6 dB . Considering both BER and average latency performance, we conclude that, among the TTs we have examined, T3.2 is the best choice for ETDs.

The numerical results presented so far assume no memory constraint. Fig. 8 and 9 reveals the impact of finite memory size for the system that employs a T3.2-aided VTT-APP decoder and the memory management algorithm of the previous section with block length $L = 400$, interleaving span $S = 1$ and $M_d = 1$. Fig. 8 shows BER performance for different memory constraints. For convenience of comparison, we also present three cases without memory constraint, one with $D_{max} = 200$, the other two with fixed DRs. It is reasonable to find that larger memory sizes give better performance. At higher $E_b/N_0 (> 0.8$ dB), all performance curves converge to the same one since all VTT-APP decoders finish decoding after only a few DRs (see also Fig. 9) and memory size is no longer a problem. The fact that the cases $D_{max} = 100$ with 100 MUs, and $D_{max} = 30$ with 100 MUs give almost identical performance indicates that increasing $D_{max}$ beyond a certain number (30 in this case) can not improve BER performance and the memory size becomes the dominant factor. Performance for the decoder with $D_{max} = 200$ and no memory constraint (it can be shown that 804 MUs is sufficient for this case, which is at least eight time larger than that required by other decoders) is clearly better than the other decoders when $E_b/N_0 < 0.6$ dB but this edge is gradually diminished after 0.6 dB.

The average DR performance is given in Fig. 9. For $E_b/N_0 \geq 0.5$ dB, all VTT-APP decoders need less than or equal to 10 DRs (5 iterations). But when $E_b/N_0 < 0.3$ dB, the performance curves are distinctly different–if we do not impose a memory constraint, the average DR will increases exponentially as $E_b/N_0$ decreases. Most of the computation effort will be wasted, so is the memory. In other words, at the low $E_b/N_0$ region, TTs can not offer early terminations. Imposing a memory constraint and invoking a proper memory management algorithm do provide a solution that forces early terminations, saving computing power and memory at the cost of a small performance loss. Finally, we find that, compared with our proposed schemes, the two decoders with fixed DRs (20 and 30) usually need much more memory and DRs.

The effectiveness of various TTs on the performance of a CTC with $L = 800$ are shown in Fig. 10 and Fig. 11. $D_{max} = 30$ DRs and tail-padding encoding are assumed. The BER performance of T1.1 improves as the length of the CRC code increases. T1.1 requires the minimum average DRs at the cost of BER performance degradation. The performance of T1.2 with CRC-8 is similar to that of T1.1 with CRC-16. T1.3 with CRC-8 already yields performance that is almost identical to that of the genie TT, so are T2.3 and T3.2. A higher-order TT leads to better decisions although the BER performance gain diminishes as $m > 2$ for all TTs considered. We also find that the average DRs for T3.2 is only one more than that required by the genie TT while their BER performance is about the same. When compared with its performance in IBPTC systems, T3.2 needs one more average DR but suffers from significant performance loss. However, if we take into account the BER and latency performance in both CTC and IBPTC systems, T3.2 is still the best candidate test. It provides better error detection capability than T1.1 with CRC-24, which confirms our conviction that a multiple-round test with a short CRC code is better than a single-round test with a much longer CRC code.

# 7  Conclusion

We present a powerful IBPTC-based coding scheme and propose the associated decoder architecture and algorithms for high speed communications. Using a multiple-round termination test and a dynamic memory management scheme, our decoder yields performance achievable by a conventional turbo coded system with higher $E_b/N_0$ and much larger decoding latency. The highly reliable TTs require only a short CRC code (low overhead) and binary sign checks. The memory management scheme makes efficient use of the storage space while maintaining low average decoding latency even at low SNR region. The decoder structure is such that expanding the decoder memory size increases the dynamic range of the memory manager or the interleaving span. Both lead to improved performance.

The new coding scheme offers a variety of design options that are not available to CTCs. It also provides increased degrees of freedom for the same design option. For the same number of ADUs, much more flexible decoding schedules are available. Its decoding is amenable for highly dynamic decoding schedules that are both distributive and cooperative: sharing all modularized decoding resources-the ADUs, interleavers/deinterleavers, memory-while passing information amongst component decoders.

System performance can be improved by using a proper decoding schedule, increasing the block size, the IBP period, the number of decoding iterations, the memory space, and the number of blocks involved in decoding. The design allow tradeoffs amongst performance, latency, computing and hardware complexities, e.g., the block size can be traded for other parameters without performance loss and lower memory requirement and higher degrees of parallelism (including memory access) are feasible by shortening the block size and using more flexible decoding resource management scheme. Multiple data sequences can be decoded in parallel and throughput is limited only by the degree of parallelism bestowed in the design.

Finally, we want to remark that, like some application layer FEC codes such as raptor code or LT code our proposal is a stream-oriented scheme. Moreover, since any existing block-wise interleaver can be regarded as an IBPI with $S = 1$, the associated APP decoders are reusable and, in a sense, our proposal is backward compatible, which makes the evolution from existing standard TCs easy and natural.

# References

[1] C. Berrou, A. Glavieux, and P. L. Thitimajshima, "Near Shannon limit error-correcting coding and decoding: turbo-codes," in *Proc. ICC'93*, pp. 1064-1070, May 1993.

[2] Y.-X. Zheng, Y. T. Su, "A new interleaver design and its application to turbo codes," *Proc. VTC2002fall*, vol.3, pp. 1437-1441, Sep. 2002.

[3] Y.-X. Zheng, Y. T. Su, "On inter-block permutation and turbo codes," in *Proc. International Symp. Turbo Codes and Related Topics*, Brest, France, Sep. 2003.

[4] J. Hagenauer, E. Offer, and L. Papke, "Iterative decoding of binary block and convolutional codes," *IEEE Trans. Inform. Theory*, vol. 42, no. 2, pp. 429-445, Mar. 1996.

[5] R. Y. Shao, S. Lin, and M. P. C. Fossorier, "Two simple stopping criteria for turbo decoding," *IEEE Trans. Commun.*, vol. 47, no. 8, pp. 1117-1120, Aug. 1999.

[6] A. Matache, S. Dolinar, F. Pollara, "Stopping rules for turbo decoders," in *TMO Progress Report 42-142*, 15 Aug. 2000.

[7] Y. T. Su and C. L. Weber, "On the performance evaluation of variable dwell time acquisition systems," in *Conf. Record, IEEE MILCOM'84*, Los Angeles, October 1984.

[8] *TS 25.222 V3.1.1 multiplexing and channel coding (TDD),* 3GPP TSG RAN WG1, Dec. 1999.

[9] G. D. Forney, Jr., "Codes on graphs: normal realizations," *IEEE Trans. on Inform. Theory*, vol. 47, pp. 520-548, Feb. 2001.

[10] Y.-X. Zheng, Y. T. Su, "On inter-block permuted turbo codes," submitted to *IEEE Trans. Inform. Theory*, see also http://arxiv.org/abs/cs.IT/0602020

[11] D. Agrawal, A. Vardy, "The turbo decoding algorithm and its phase trajectories," *IEEE Trans. Inform. Theory*, vol. 47, no. 2, pp. 699-722, Feb. 2001.

[12] M. Luby, "LT codes" in *Pro. 43rd Annual IEEE Symp. Foundations of Computer Science*, pp. 271-280, 2002.

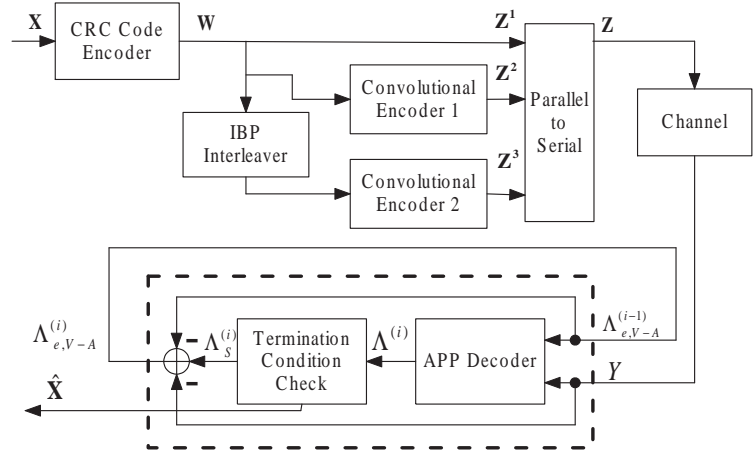

Figure 1: A block diagram for the proposed coding system in which the encoder uses an IBP interleaver. More than one VTT-APP decoding unit can be used; the notations denote various extrinsic information for the $i$th unit.

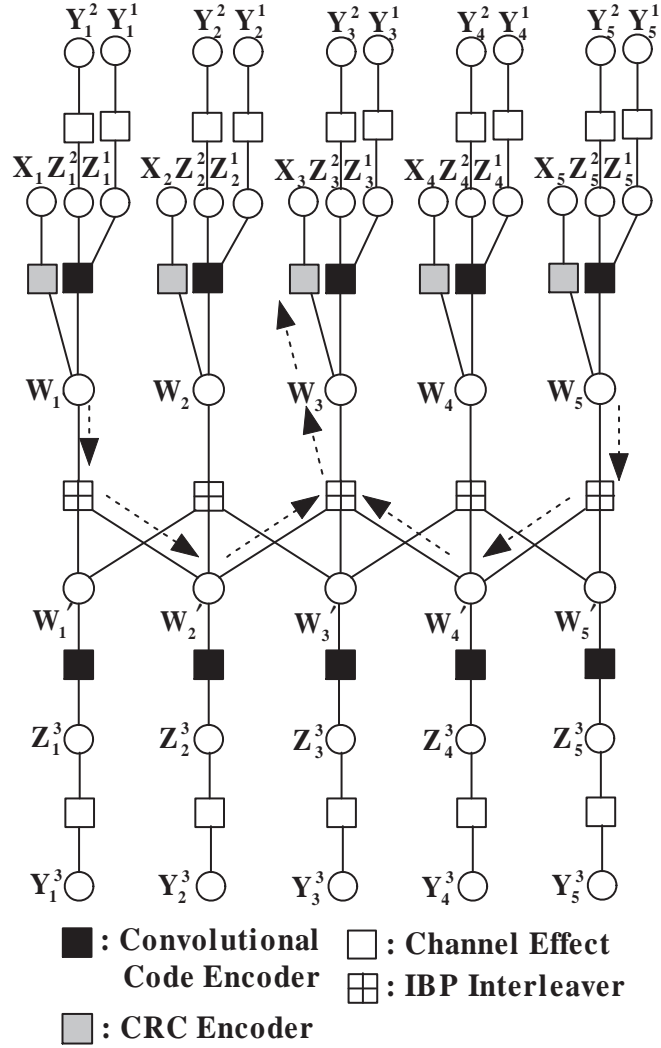

Figure 2: A factor graph representation for a CRC and IBPTC encoded system with interleaving span $S = 1$.

|  CTC/IBPTC | | | | | | | |
|---|---|---|---|---|---|---|---|
| APP Decoding Round \ Block Index | 1 | 2 | 3 | 4 | 5 | 6 | 7 |
| 1 | 1/1 | 5/2 | 9/4 | 13/7 | 17/11 | 21/15 | 25/19 |
| 2 | 2/3 | 6/5 | 10/8 | 14/12 | 18/16 | 22/20 | 26/23 |
| 3 | 3/6 | 7/9 | 11/13 | 15/17 | 19/21 | 23/24 | 27/26 |
| 4 | 4/10 | 8/14 | 12/18 | 16/22 | 20/25 | 24/27 | 28/28 |

Figure 3: A comparison of exemplary decoding schedules for CTC and IBPTC when decoding 7 blocks with 2 iterations (four DRs). The numbers $a/b$ in the constituent squares represent the order the APP decoder performs decoding for CTC/IBPTC. Hence the first block of the CTC is decoded by the first 4 DRs (the left upper numbers in the second leftmost column) but that of the IBPTC is decoded by the first, third, sixth and tenth DRs (the right lower numbers in the same column); see subsection 2.3 for detailed discussion.

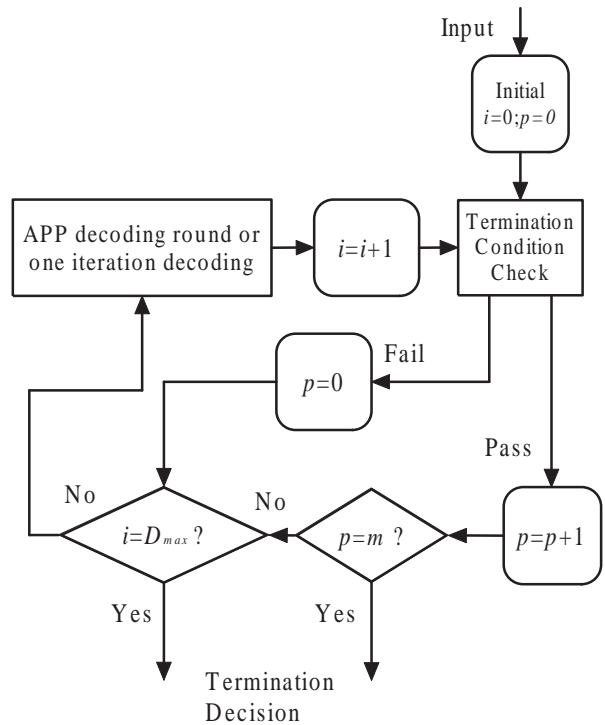

Figure 4: Flow chart of a general $m$-round termination test (MRTT).

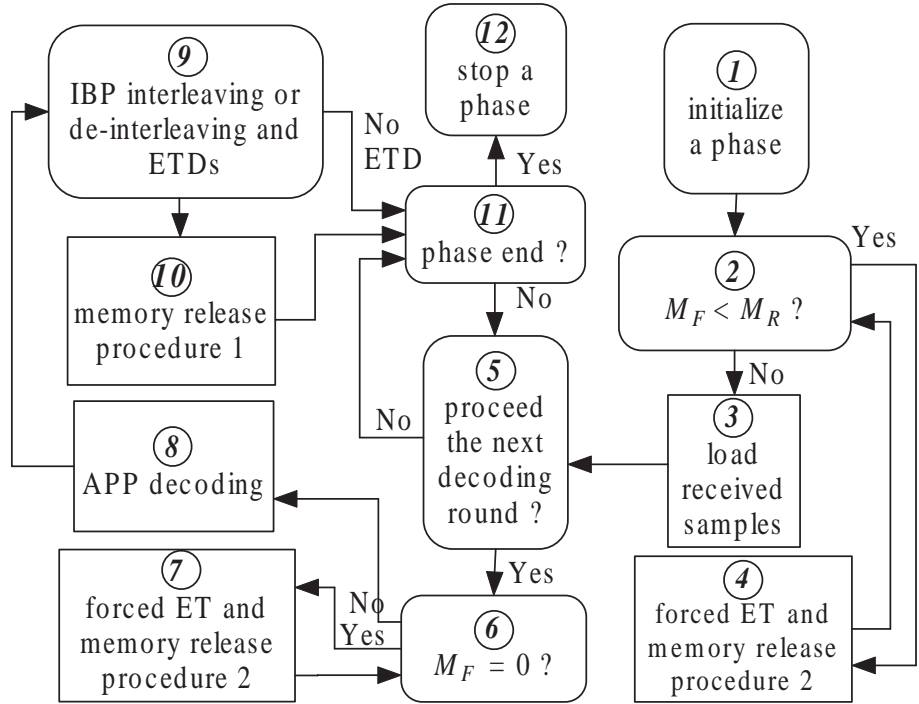

Figure 5: (a) A multiple zigzag decoding schedule for IBPTCs with an interleaving span $S = 1$; (b) A joint memory management and IBPTC decoding procedure.

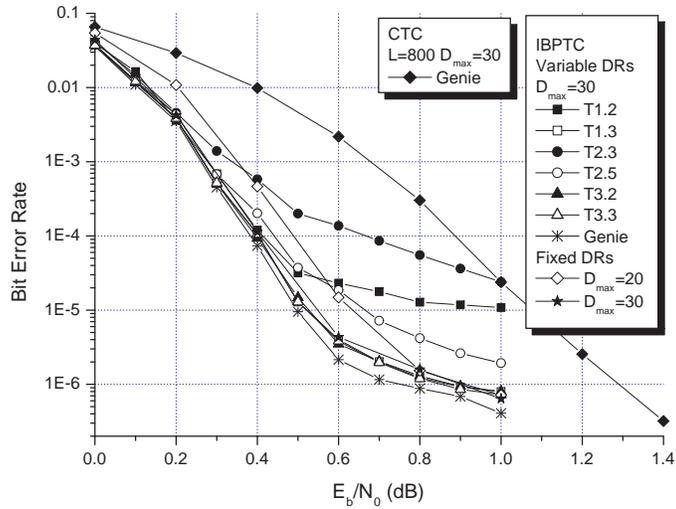

Figure 6: Bit error rate performance of various termination tests; no memory constraint; $D_{max} = 30$ DRs.

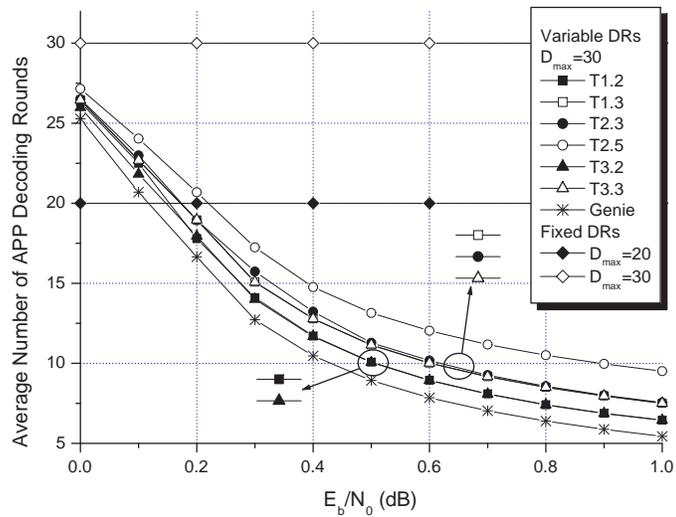

Figure 7: Average APP DR performance of various termination tests; $D_{max} = 30$ DRs, no memory constraint.

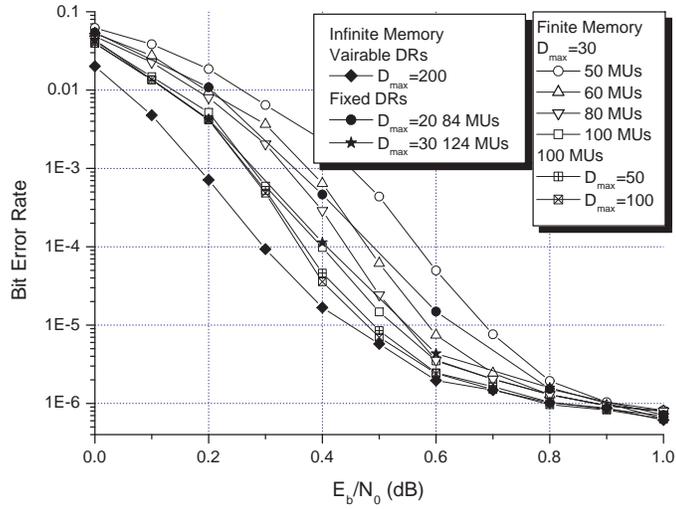

Figure 8: The effect of memory constraint and management on the bit error rate performance. Curves labelled with infinite memory are obtained by assuming no memory constraint; fixed DRs implies that no early-termination test is involved.

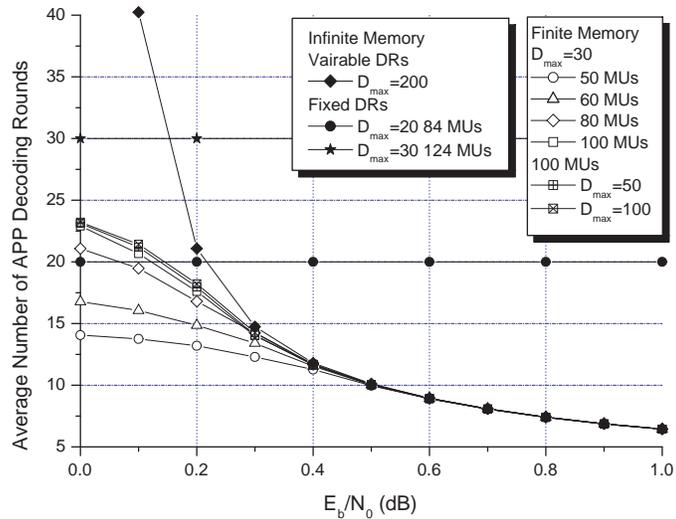

Figure 9: Average APP DR performance for various decoding schemes and conditions. Curves labelled with infinite memory are obtained by assuming no memory constraint; fixed DRs means no early-termination condition is imposed.

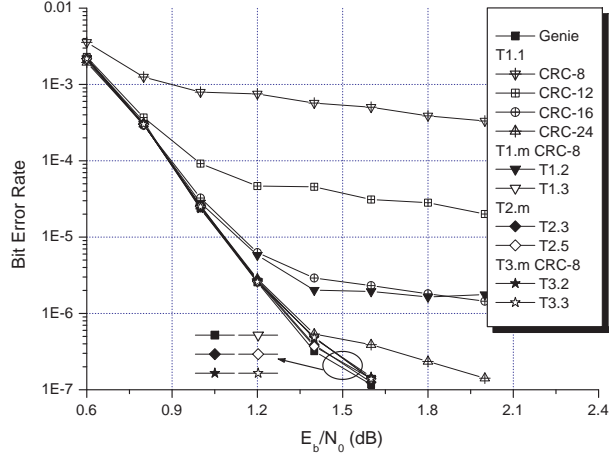

Figure 10: Bit error rate performance of a CTC using various TTs; $L = 800$ bits and $D_{max} = 30$ DRs.

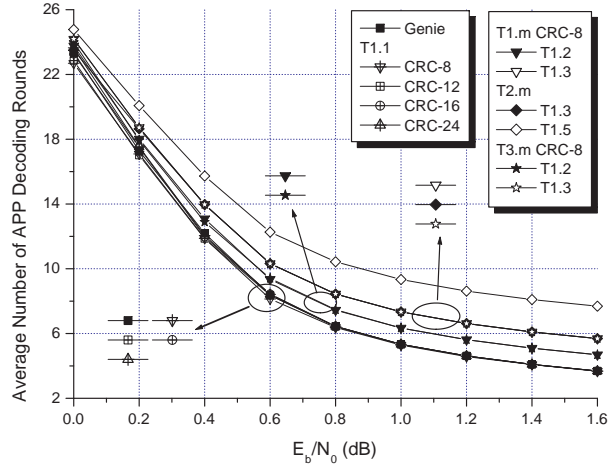

Figure 11: The effect of Various TTs on the average APP DR performance of a CTC with $L = 800$ and $D_{max} = 30$ DRs.

Full memory usage = $D_{max}S(M_R + 1) + M_d \cdot (M_R + 1)$

2 Memory check condition $M_F < M_R + M_d$: Allocate $M_R$ MUs for received samples and $M_d$ MUs for $M_d$ parallel processors.

7 Memory check condition $M_F < M_d$: Allocate $M_d$ MUs for $M_d$ parallel processors.

6 Proceed to the $n$th decoding ? : Decision make by previous termination test. If "Yes", perform the following APP decoding, IBP permutation, termination test and memory release procedure. If "No", perform memory release procedure only.

12 $n = D_{max}$? checks if maximum decoding run reached.

14 Memory release ? : Proceed on odd decoding run.

10 IBP interleaver and deinterleaving depend on APP decoding corresponding to pre-permutation and post-permutation. Termination test proceeds on tentative decoded stream corresponding to pre-permutation stream.

11 For memory release procedure 1, we consider three parts: 1) received samples including systematic part and parity check part corresponding to pre-permutation; 2) received samples of parity check part corresponding to post-permutation; 3) tentative decoded hard bits.

Part 1 is released only when a block passes the TT. If pass termination test, tentative decoded result will be record by a stream of bits and generate extrinsic information as "$\pm\infty$" if required.

Part 2 of the $s$th block before the current one will be released if the $2s$ blocks before the current one have also pass TT. A pass flag will be assigned to that block as well.

Part 3 the $2s$th block before the current one shall be released if all previous $4S$ blocks pass TT.

4,8 Memory release procedure 2

odd: makes hard-decision, output data, release all information, including received samples, extrinsic information and decoded result, corresponding to the "oldest" block. The other blocks within its span will be semi-terminated, in the sense that no more APP decoding will be performed except for interleaving/deinterleaving.

> even: makes HD on the "corresponding" blocks .....